# Multiband Superconductivity and Charge Density Wave in $Hf_xZr_{1-x}Te_3$ single crystals

F. F. G. Nogueira[a, α], M.S. da Luz [b], M.S. Torikachvili [c], L.M. Ishikura[a], Ebrahim Dib[a], A.J.S. Machado[a, β]

[a] Departamento de Engenharia de Materiais, Universidade de São Paulo, Escola de Engenharia de Lorena, Lorena, SP 12612-550, Brazil
[b] Instituto de Ciências Tecnológicas e Exatas, Universidade Federal do Triangulo Mineiro, Uberaba, MG 38025-180, Brazil
[c] Department of Physics, San Diego State University, San Diego, CA 92182, USA

[α]Contact author: fanogueira@usp.br

[β]Contact author: ajefferson@usp.br

**Abstract**

We report on the bulk multiband superconductivity and charge density wave (CDW) in $Hf_xZr_{1-x}Te_3$ single crystals. The parent compound $ZrTe_3$ is a layered van der Waals material that undergoes a CDW transition near 63 K and exhibits only filamentary superconductivity below 2.0 K. Upon the introduction of a small amount of Hf (x = 0.02), the CDW transition temperature is reduced to $T_{CDW} \approx 53$ K, while a robust bulk superconducting state emerges at $T_c \approx 3.3$ K, underscoring a subtle competition between CDW order and superconductivity in this quasi-one-dimensional system. Electrical resistivity, magnetic susceptibility, Hall effect, Seebeck coefficient, and specific heat measurements consistently confirm the bulk nature of the superconducting phase. The temperature dependence of the upper critical field $\mu_0 H_{c2}(T)$ deviates markedly from single-band behavior, and it is well described by a two-band model, consistent with multiband superconductivity. Analysis of the Hall effect, and thermoelectric behavior reveal pronounced electronic anisotropy, with enhanced effective carrier masses, indicating that the subtle structural modification introduced by Hf substitutions affects the Fermi surface topology, as well as electronic correlations. Measurements of electrical resistivity in hydrostatic pressures up to ≈ 2 GPa reveal that pressure drives $T_{CDW}$ to higher temperatures while suppressing $T_c$. These findings show that Hf doping can be used to fine-tune the balance between the CDW instability and superconductivity, possibly by means of chemical pressure effects, stabilizing a multiband superconducting state in Hf-doped $ZrTe_3$.

**Introduction:**

The study of competing and coexisting collective electronic phases is of fundamental interest in condensed matter physics [1]. Among these phenomena, the interplay between charge density wave (CDW) and superconductivity (SC) has attracted particular interest because both states partially gap the Fermi surface and yet arise from distinct symmetry-breaking mechanisms. CDW order breaks translational symmetry, typically driven by Fermi surface nesting and/or strong electron–phonon coupling, whereas conventional superconductivity emerges from phonon-mediated Cooper pairing. Importantly, both CDW and SC originate from intrinsic electronic instabilities that are especially pronounced in low-dimensional systems [2, 3, 4]. In such materials, enhanced nesting conditions and strong electron–phonon interactions favor CDW formation, establishing a delicate balance in which relatively small perturbations, such as chemical substitution, disorder, or external pressure, can suppress the CDW state and perhaps stabilize superconductivity instead [1, 5, 6].

Layered transition metal trichalcogenides (TMTs) constitute a prototypical family of low-dimensional materials quite amenable for probing the competition between CDW and SC. These compounds have general composition $MX_3$ (M = Zr, Hf, Ti, Nb; X = S, Se, Te) and exhibit quasi-one-dimensional (quasi-1D) or strongly anisotropic electronic structures, leading to complex phase diagrams where CDW, SC, and other electronic orders compete or coexist [3, 4, 6, 7]. A prominent example is $ZrTe_3$, a van der Waals uniaxial compound that crystallizes in the $ZrSe_3$-type structure (space group $P12_1/m_1$) [8, 9]. Its structure consists of infinite quasi-1D chains of Te trigonal prisms running along the crystallographic b-axis, each with a Zr ion at its center. These chains are weakly coupled along the c-axis via van der Waals interactions, forming a monoclinic lattice [8, 9, 10]. The natural cleavage surface corresponds to the ab plane, and the longest crystal dimension lies typically along the b-axis, reflecting the direction of strongest covalent bonding; large crystal facets frequently display fine striations along this direction [11]. Pure $ZrTe_3$ undergoes a CDW transition at $T_{CDW} \approx 63$ K, primarily associated with an electronic instability along the a-axis. Filamentary superconductivity with $T_c \approx 2$ K under ambient conditions can only be observed for excitation current along the b-axis [9, 12, 13, 37].

Several external and chemical tuning strategies have been shown to suppress the CDW instability in $ZrTe_3$ and promote bulk superconductivity. These include intercalation with transition metals (Ni, Cu, Ru) or rare-earth elements (Tb, Er), substitutional doping (e.g., Se at the Te site), disorder, and hydrostatic pressure, all revealing that the CDW–SC interplay in $ZrTe_3$ is highly tunable [5, 11, 14-19]. In some cases, the behavior of the upper critical field $H_{c2}$ is consistent with multiband superconductivity [11, 16, 17]. In addition, previous studies have reported a disorder-related SC transition with $T_c \approx 4.3$ K, and a CDW formation with $T_{CDW} \approx 93$ K in $HfTe_3$ single crystals [20, 21].

In this study we probed the competition of the CDW and SC in $Hf_xZr_{1-x}Te_3$ single crystals of two compositions, x = 0.02 and 0.13, by means of measurements of electrical, thermal, and magnetic properties. The value of $T_c$ determined from the onset of the superconducting transition in the resistivity data ρ(T) is ≈ 3.3 K; the magnetization data below $T_c$ display clear type-II behavior, and the behavior of $H_{c2}$ vs T extracted from the ρ(T) data is well described by a two-band model. Superconductivity is suppressed and the CDW state becomes more pronounced for higher concentrations of Hf. A similar trend of strengthening the CDW while suppressing SC is observed under hydrostatic pressure.

**Materials and Methods:**

Single crystals of $Hf_xZr_{1-x}Te_3$ were grown from high-quality polycrystalline precursor pellets with nominal composition $Hf_xZr_{1-x}Te_3$ (x = 0.1, 0.2, and 0.3) using an isothermal chemical vapor transport (ICVT) method [22]. The precursors were synthesized by reacting stoichiometric amounts of high-purity Hf (99.7%, Aldrich), Zr (99.5%, Alfa Aesar), and Te (99.999%, Alfa Aesar) powders at 900 °C for 48 h in evacuated and sealed quartz tubes, followed by rapid quenching in cold water. The resulting materials (0.5 g batches) were thoroughly ground and cold-pressed into pellets with 8 mm diameter. An additional annealing cycle under identical conditions was performed to ensure compositional homogeneity.

For the growth process, each precursor pellet was loaded in a quartz ampoule together with 15 mg of $I_2$ (99.99%, Alfa Aesar), which served as the transport agent, evacuated to ≤ 100 mtorr pressure, and sealed. The ICVT process was conducted at 900 °C for 7 days, followed by furnace cooling to room temperature. The crystals that grew out of the pellets were thin needle- and tape-shaped with the large facets along the ab-plane, and elongated along the b-axis, with largest dimensions ≈ 6.0 × 0.8 × 0.15 mm³.

The crystal structure and phase purity were examined by X-ray diffraction (XRD) using $CuK_\alpha$ radiation (λ = 1.5418 Å) on a Panalytical Empyrean diffractometer. The XRD patterns confirmed phase purity and excellent crystallinity. As expected for the low substitution level (~2%), no significant variation in lattice parameters was detected. The chemical composition of the $Hf_xZr_{1-x}Te_3$ crystals was determined by energy-dispersive spectroscopy (EDS) using a Hitachi TM-3000 scanning electron microscope.

Measurements of the electronic, magnetic and thermal properties were carried out using the options of a Quantum Design Physical Property Measurement System (PPMS)-EverCool II. Electrical resistivity measurements were performed using the standard four-probe configuration. The samples were first glued on a small section of Kapton tape with GE-7031 varnish, after which four 50 µm dia. copper wires were attached using nickel-loaded epoxy. Resistivity was measured along both the a-axis and the b-axis. Measurements of resistivity in pressures up to ≈ 2 GPa were conducted using a piston–cylinder self-locking pressure cell, adapted for measurements with the PPMS. The sample, a thin strip of Pb, and a coil of manganin, which served as low- and high-temperature manometers, respectively, were attached to the end of a Ni-Cr-Al feedthrough using a 4-wire configuration. The feedthrough was inserted in a PTFE cup filled with a pressure transmitting medium (40:60 mixture of light mineral oil and n-pentane). Force was applied and locked in at ambient temperature with a hydraulic press. The actual pressure in this type of cell is known to change with temperature before stabilizing near 90 K [36]. Pressure values between 90 and 300 K were estimated from a linear interpolation of the values yielded by the Pb and manganin manometers.

The Hall coefficient was determined using a 4-point configuration. The samples were placed with the ab-plane perpendicular to the magnetic field. The electrical current was applied at the end of the samples, either along the a- or b- directions, and the Hall voltages were detected by two wires located on the sides, across from each other. In order to subtract the longitudinal component due to the possible misalignment of the voltage leads, the Hall voltage was measured with both orientations of the field, such that the longitudinal voltage could be cancelled by antisymmetrization. The Hall resistance was taken from $R = (R_+ - R_-)/2$, where $R_+$ and $R_-$ represent both orientation of the field, and the Hall coefficient was calculated from $R_H = R.t/B$, where t is the thickness of the sample and B is the field magnitude.

The quoted values of $T_{c,\ onset}$ taken from the resistivity measurements correspond to 10% drops of the $\rho(T)$ from the normal values, while the values of $T_{CDW}$ were taken from the minimum in the derivatives $d\rho_a(T)/dT$.

**Experimental Results and discussion:**

The crystal structure of the $Hf_{0.02}Zr_{0.98}Te_3$ crystals was confirmed by X-ray diffraction, as shown in Fig. 1. For the $\theta-2\theta$ scan of Fig. 1, the sample was laid flat on the ab-plane. All reflections can be indexed to the ($00\ell$) family of planes of the parent compound $ZrTe_3$, which crystallizes in the space group $P12_1/m_1$. The sharp reflections indicate high crystallinity and phase purity of the crystal [8, 9].

The inclusion of Hf resulted in a minor reduction of the lattice parameter *c*, consistent with Vegard's law. However, given the proximity of lattice parameters of $ZrTe_3$ and $HfTe_3$, where a-, b-, and c- are reduced by ≈ 0.32, 0.63, and 0.47%, respectively, it is difficult to determine precisely how much of the Hf in the samples is substitutional or intercalated between the layers.

The incorporation of Hf in the single crystals is confirmed by EDS. The EDS analysis of the crystals grown from the $Hf_{0.1}Zr_{0.9}Te_3$ precursor pellet yielded the $Hf_{0.02}Zr_{0.98}Te_3$ composition. The $Hf_{0.2}Zr_{0.8}Te_3$ and $Hf_{0.3}Zr_{0.7}Te_3$ precursors yielded crystals with composition $Hf_{0.13}Zr_{0.87}Te_3$ and $Hf_{0.25}Zr_{0.75}Te_3$, respectively.

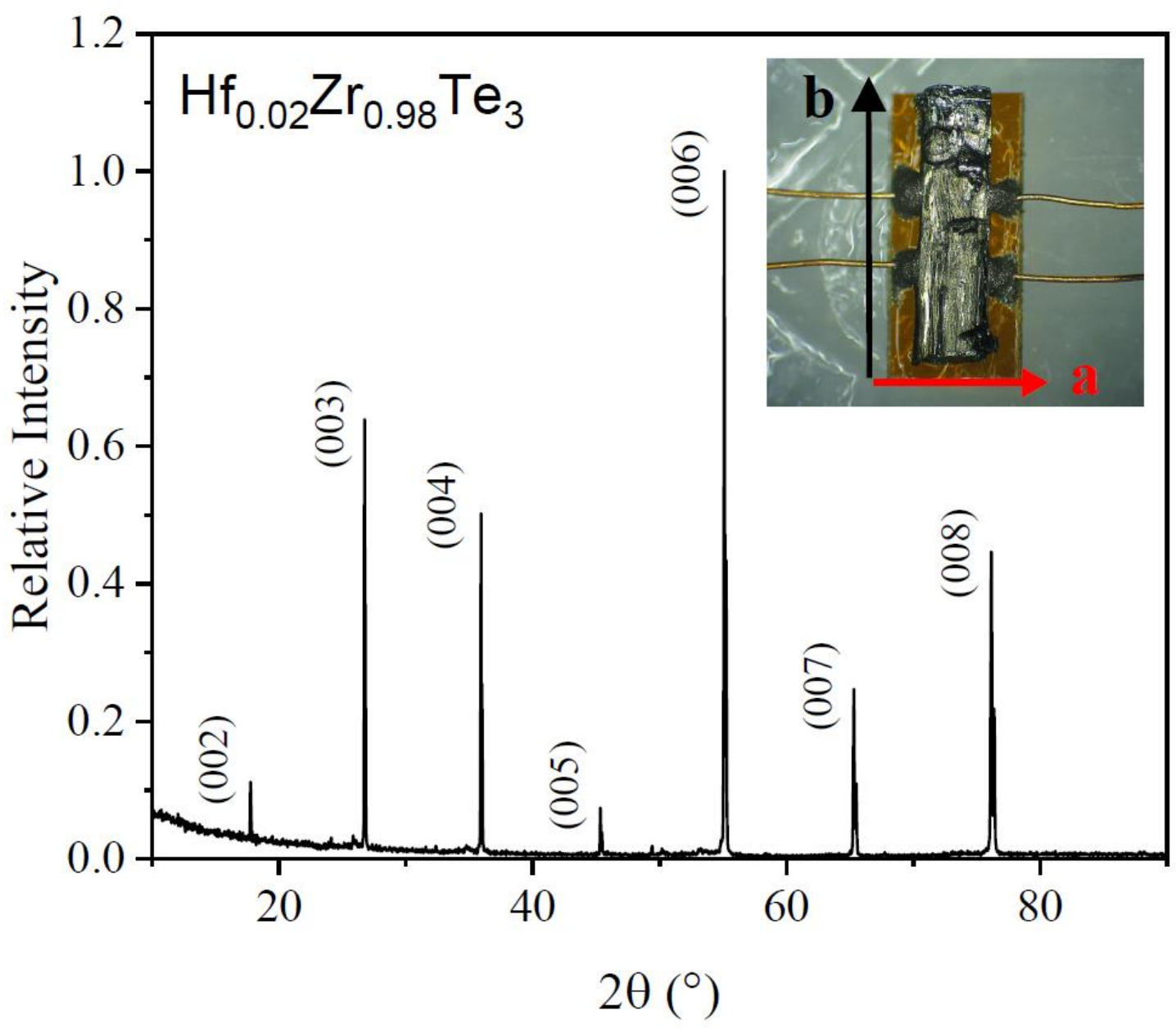

**Figure 1.** θ-2θ X-ray diffraction pattern for the $Hf_{0.02}Zr_{0.98}Te_3$ single-crystal showing only the (00$\ell$) reflections. The inset shows the $Hf_{0.02}Zr_{0.98}Te_3$ crystal used for the electrical resistivity measurement with excitation current along the a-axis.

The temperature dependence of the electrical resistivity along the *a*-axis $\rho_a(T)$ for the $Hf_xZr_{1-x}Te_3$ single crystals with x = 0.02, 0.13, and 0.25 is shown in Figure 2. Superconductivity with a $T_{c,\ onset} \approx 3.3$ K emerges clearly for the lightly substituted samples with x = 0.02 and 0.13 as shown in the inset Fig. 2(a), displaying $\rho_a(T)/\rho_{a,\ 5K}$. No indication of drop in resistance down to 1.8 K was observed in the x = 0.25 sample.

A pronounced anomaly associated with the CDW transition in the 50 - 70 K temperature range is clearly visible in the $\rho_a(T)$ data for the three compositions, as show in the inset Fig. 2b. Taking $T_{CDW}$ from the minimum in $d\rho_a(T)/dT$, we can infer that the suppression of superconductivity at larger Hf concentrations correlates well with the increase in $T_{CDW}$, as shown in Fig. 2b inset. $T_{CDW}$ increases systematically with Hf content, rising from ≈ 53 K for x = 0.02 to ≈ 72 K for x = 0.25. Thus, as the CDW order strengthens with increasing Hf substitution, superconductivity is progressively suppressed, suggesting a competitive relationship between these two collective electronic orders.

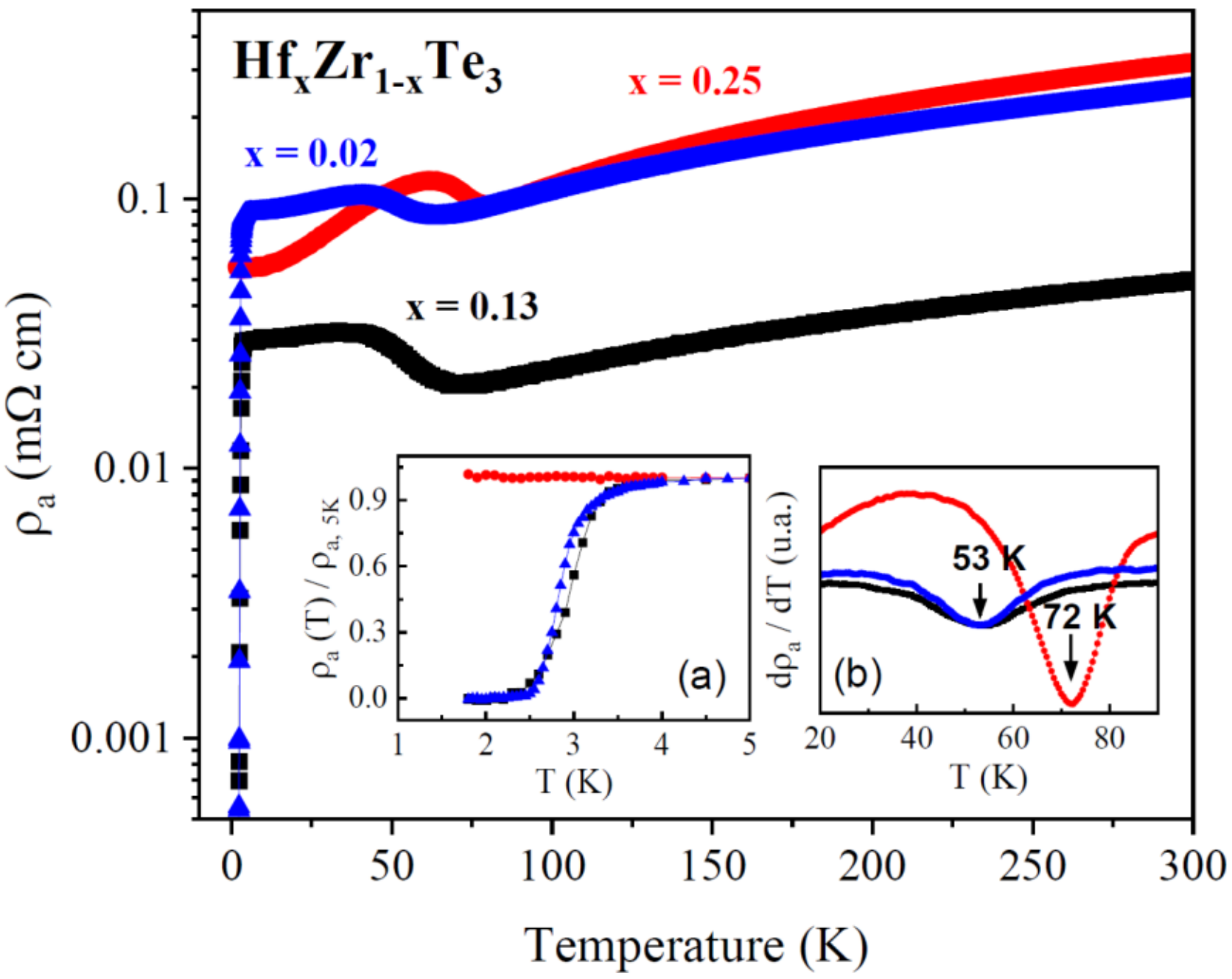

**Figure 2.** Temperature dependence of the electrical resistivity along the *a*-axis, $\rho_a(T)$, for $Hf_xZr_{1-x}Te_3$ single crystals with Hf concentrations x = 0.02 (blue), 0.13 (black), and 0.25 (red), plotted on a logarithmic scale. Insets: (a) Low-temperature region showing the normalized resistivity $\rho_a(T)/\rho_{a,\,5\,K}$; (b) Temperature derivative of the resistivity, $d\rho_a/dT$, highlighting the charge density wave transition. The CDW transition temperature, $T_{CDW}$, was determined from the minimum in $d\rho_a/dT$.

Albeit disorder, pressure, or carrier doping more commonly reduce the value of $T_{CDW}$, [5, 11, 14-19], the $Hf_xZr_{1-x}Te_3$ compounds are unique in which the $T_{CDW}$ value increases with the partial substitution of Hf for Zr. Here, isovalent partial substitution does not significantly alter the carrier concentration, or lattice parameters. The observed enhancement of $T_{CDW}$ suggests improved nesting conditions or strengthened electron–phonon coupling within the quasi-one-dimensional electronic states. As the CDW becomes more robust, a larger portion of the Fermi surface is expected to be partially gapped, therefore reducing the density of states available for superconducting pairing and ultimately suppressing the onset of superconductivity.

It is important to notice that the metallic behavior of $\rho_a(T)$ above $T_{CDW}$ is preserved in the composition range (x = 0.02-0.25), indicating that the itinerant carriers are not affected much by disorder or the presence of strong electronic correlations. The strengthening of the CDW upon doping without an increase in disorder suggests that the tuning mechanism is predominantly electronic rather than impurity-driven.

The absence of superconductivity at high Hf concentrations does not contradict previous reports of filamentary superconductivity in $HfTe_3$ crystals, where filamentary disorder-related superconductivity was observed at $T_c \approx 4.3$ K [20, 21]. In $ZrTe_3$-type compounds, filamentary and bulk superconductivity originate from distinct electronic subsystems: the filamentary state develops on quasi-one-dimensional (quasi-1D) Fermi-surface sheets associated with the CDW instability, whereas bulk superconductivity emerges from a more three-dimensional (3D) electronic component capable of sustaining long-range phase coherence [9, 10, 34]. The $\rho_a(T)$ data indicate that increasing Hf substitution progressively drives the system toward a more robust CDW regime. As the CDW strengthens, it is tempting to consider that a larger fraction of the quasi-1D Fermi surface becomes gapped, reducing the electronic density of states available for superconducting pairing.

Within this framework, high Hf concentrations suppress the 3D electronic component required to stabilize a bulk superconducting phase, even if short-range or filamentary superconducting correlations confined to residual quasi-1D segments may persist. It is plausible to consider that Hf substitution drive a gradual reduction of the effective electronic dimensionality, shifting the balance from a regime where bulk superconductivity can coexist with a weakened CDW to one in which the CDW dominates, and long-range superconducting coherence is energetically disfavored. While direct momentum-resolved probes of the Fermi surface evolution in Hf substituted composition would be required to evaluate this scenario, the systematic suppression of bulk superconductivity with Hf content provides credence to this interpretation.

Within the context of the coexistence and competition of superconductivity and CDW, it is instructive to contrast the $\rho_a(T)$ data for the Hf partially substituted $ZrTe_3$ compounds with those reported for Tb intercalation, whose incorporation enhances bulk superconductivity while simultaneously suppressing the CDW state [11]. It's been reported that filamentary superconductivity in $ZrTe_3$ has a $T_{c,\ onset} \approx 2.0$ K, and a CDW order with $T_{CDW} \approx 63$ K [9, 12, 13, 37]. However, $T_{c,\ onset}$ reaches $\approx 5.4$ K in $Tb_{0.02}ZrTe_3$, while $T_{CDW}$ is reduced to $\approx 50$ K, clearly indicating that the weakening of the CDW instability corresponds to the enhancement of superconductivity. In contrast, the $\rho_a(T)$ data for the Hf-substituted $ZrTe_3$ reveal the opposite trend. It's tempting to correlate the "fine tuning" of the superconductivity and CDW ground states with the size of the substituting species. While 12-coordinated Hf ions are $\approx 1.3\%$ smaller than Zr, Tb ions are $\approx 11.3\%$ larger, therefore introducing more disorder. In addition, due to the smaller size of the substituting Hf ions, it causes a positive chemical pressure in the crystal lattice that adds up to the hydrostatic pressure, which is expected to compress even more the structure, introducing disorder and affecting the electronic structure.

Having established the presence of both CDW and superconductivity in the lightly substituted compounds, it is essential to examine how these collective states form along different crystallographic directions. The highly anisotropic crystal structure is a typical characteristic of many low-dimensional and van der Waals materials, and $Hf_xZr_{1-x}Te_3$ is no exception [9, 10, 19]. In this system, both the superconducting and CDW states exhibit strong anisotropy, reflecting the underlying quasi-one-dimensional electronic structure.

As shown in Fig. 3, the CDW transition in $Hf_{0.02}Zr_{0.98}Te_3$ appears as a clear anomaly in $\rho_a(T)$ centered near 53 K, while no corresponding feature is observed in $\rho_b(T)$. This directional selectivity indicates that the CDW instability predominantly affects electronic states associated with the *a*-axis. Although both resistivity curves display metallic behavior for $T>T_{CDW}$ ($d\rho/dT > 0$), their magnitudes differ substantially, with $\rho_a$ approximately one order of magnitude larger than $\rho_b$. This pronounced anisotropy can be quantified by analyzing the residual resistivity ratios (RRR = ρ(300 K)/ρ(6 K)), resulting in $RRR_a = 2.9$ and $RRR_b = 15.5$, corresponding to stronger metallic behavior along the *b*-axis. The inset of Fig. 3 (a) shows sharp superconducting transitions for excitation currents along the a- or b-axis, consistently with the high crystalline quality of the samples.

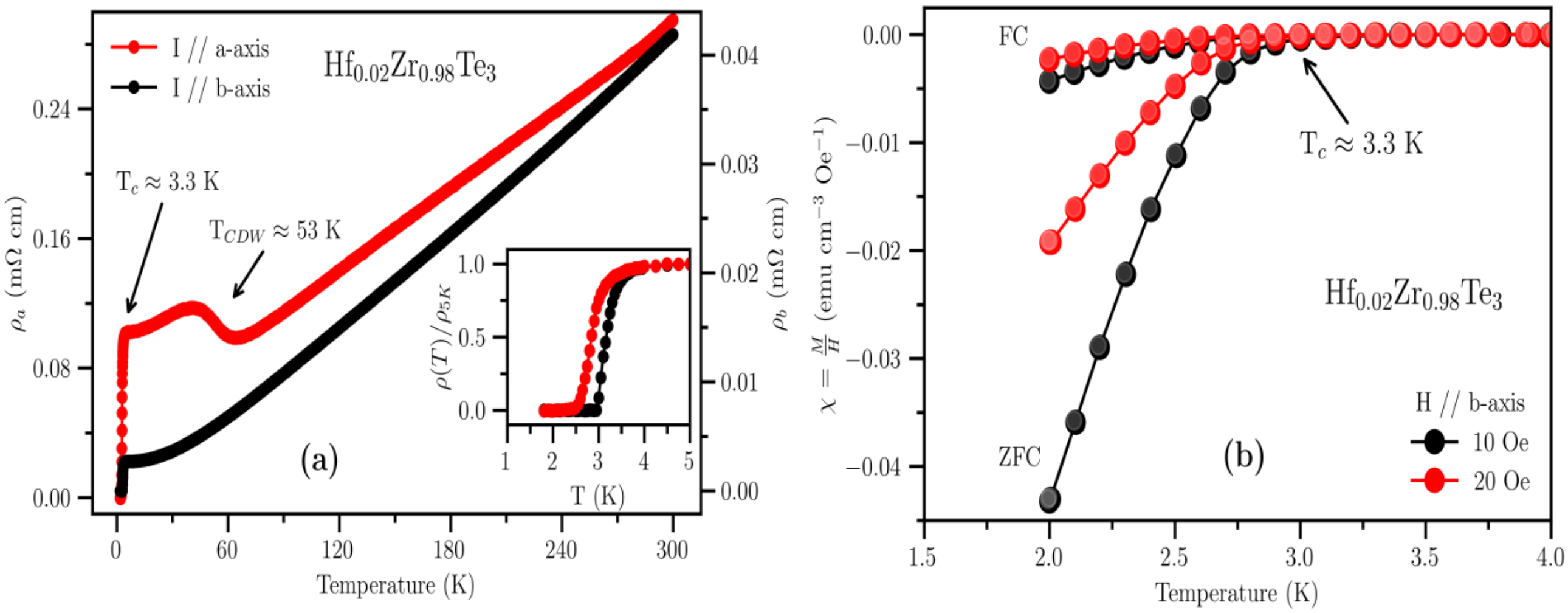

**Figure 3.** (a) Temperature dependence of the electrical resistivity for $Hf_{0.02}Zr_{0.98}Te_3$ $\rho_a(T)$ (left y-axis), and $\rho_b(T)$ (right y-axis) over the temperature range 2–300 K. The CDW anomaly at $T_{CDW} \approx 53$ K and $T_{c,\,onset} \approx 3.3$ K are indicated by arrows. The inset displays the normalized resistivity ρ(T)/ρ(5 K) in the low-temperature region (2–5 K), highlighting the superconducting transitions for both current directions; (b) Temperature dependence of the magnetic susceptibility in applied magnetic fields of 10 Oe (black dots), and 20 Oe (red dots). The divergence between the ZFC (zero-field-cooled) and FC (field-cooled) curves yields a $T_c \approx 3.3$ K for the 10 Oe applied magnetic field H // b-axis.

Band structure calculations for $ZrTe_3$-type materials attribute the strong anisotropic behavior to homonuclear Te chains running along the *a*-axis [10], which shape the quasi-one-dimensional Fermi surface sheets responsible for the CDW instability. The value of $T_c$ yielded from the magnetization measurements (Fig. 3b) correlates well with the value obtained from resistivity. The zero-field-cooled (ZFC) and field-cooled (FC) χ vs T curves for H // b-axis exhibit a clear onset of diamagnetism at $T_c \approx 3.3$ K for an applied magnetic field of 10 Oe. The drop in magnetic susceptibility and the well-defined diamagnetic signal confirm the bulk nature of superconductivity in $Hf_{0.02}Zr_{0.98}Te_3$. The small separation between ZFC and FC curves suggest weak flux pinning and good sample homogeneity, providing additional evidence that the superconducting behavior is intrinsic and bulk.

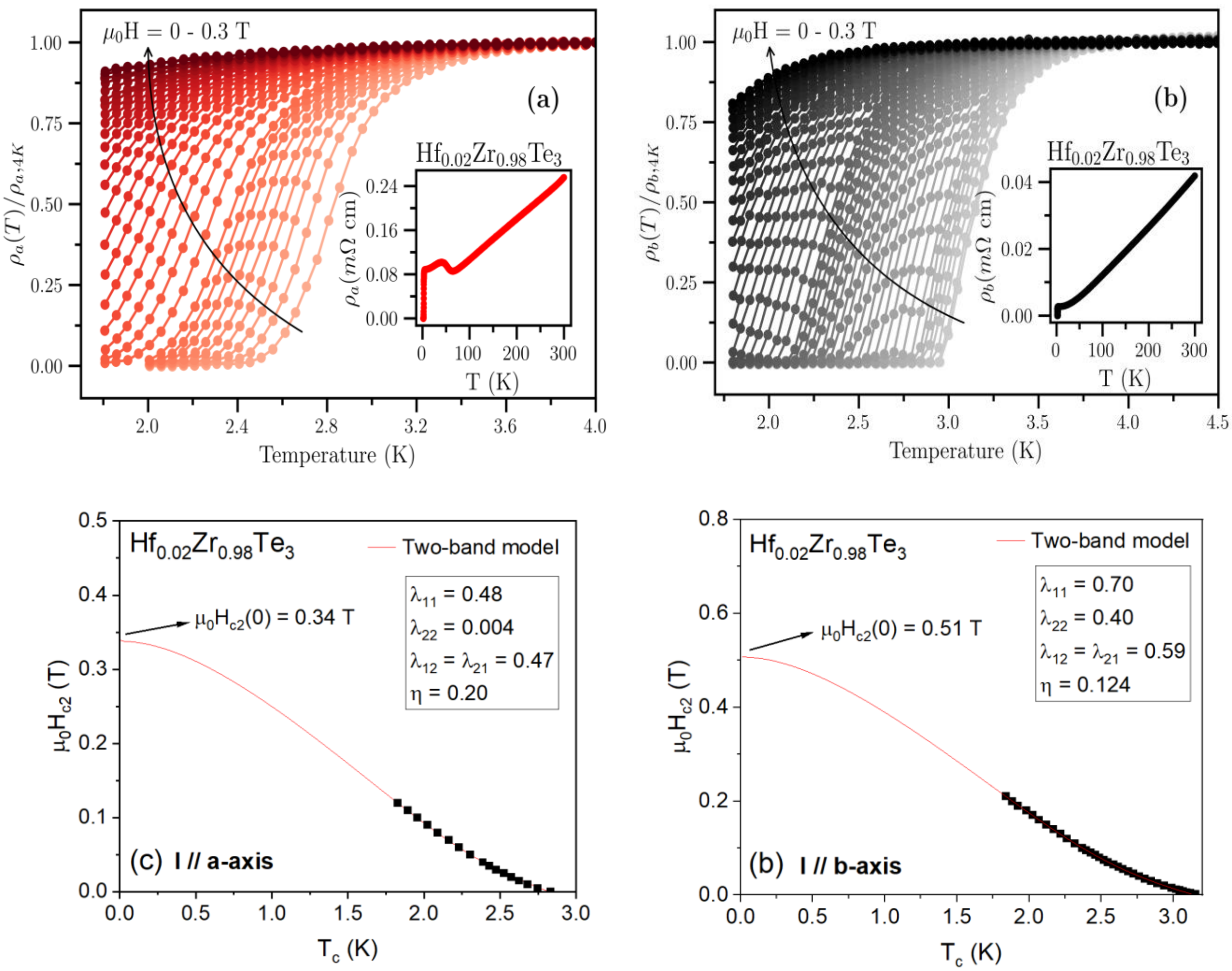

**Figure 4.** Temperature dependence of the normalized resistivity ρ(T)/ρ(4 K) for $Hf_{0.02}Zr_{0.98}Te_3$ in magnetic fields $\mu_0H$ = 0–0.3 T for (a) I // a and (b) I // b. The superconducting transition is progressively suppressed with increasing field. A small difference of the data for the two orientations of the field can be attributed to slight difference in the broadening. Insets show ρ(T) from 1.8 to 300 K. The upper critical field $\mu_0H_{c2}$ was determined using a 50% criterion for the resistive transition.

Consistent with the ρ(T) and χ(T) data of Figure 3 for $Hf_{0.02}Zr_{0.98}Te_3$, the corresponding behaviors for $Hf_{0.13}Zr_{0.87}Te_3$ are quite similar. To streamline the discussion and avoid redundance, we'll concentrate heretofore on the detailed analysis of the $Hf_{0.02}Zr_{0.98}Te_3$ single crystal data. As shown in Figure 4, increasing the applied magnetic field up to 0.3 T gradually weakens the superconducting state, causing the transition to shift systematically toward lower temperatures for both current directions, I // a (Fig. 4a), and I // b (Fig. 4b). The zero-resistance state remains observable up to approximately 70 mT for I // a, and 150 mT for I // b, indicating that superconductivity is more resilient for magnetic fields when the excitation current is along the b-axis. The different zero-resistance temperatures reflect the anisotropy. In order to carry out an analysis of the temperature dependence of upper critical field, the $\mu_0H_{c2}(T)$ values were taken from the 50% drops of the normal state resistivity at the various magnetic fields, yielding $\mu_0 H_{c2}$ vs $T_c$ superconducting phase boundaries for $Hf_{0.02}Zr_{0.98}Te_3$ as indicated in Figs. 4c and 4d.

The $\mu_0 H_{c2}(T)$ curves exhibit a pronounced positive curvature near $T_c$ for both orientations of the excitation current. This detail is particularly relevant because it cannot be explained within a conventional single-band description, which typically predicts a nearly linear or slightly downward curvature behavior close to $T_c$. Instead, the observed upward curvature indicates multiband superconductivity. The $\mu_0 H_{c2}(T)$ data can be consistently described using a two-band model [23], in which $\mu_0 H_{c2}(T)$ is obtained from the implicit relation

$$a_0[\ln t + U(h)][\ln t + U(\eta h)] + a_1[\ln t + U(h)] + a_2[\ln t + U(\eta h)] = 0 \quad (1)$$

where $U(x) = \Psi(1/2 + x) - \Psi(1/2)$, $\Psi(x)$ is the digamma function, $t = T/T_c$, $\eta = D_2/D_1$ is the ratio between the electronic diffusivities of bands 2 and 1, $h = H_{c2}D_1/2\phi_0 T$, and $\phi_0 = 2.07 \times 10^{-15}$ Wb is the magnetic flux quantum. The coefficients $a_0$, $a_1$, and $a_2$ depend on the intraband ($\lambda_{11}$, $\lambda_{22}$) and interband ($\lambda_{12} \approx \lambda_{21}$) coupling constants, reflecting the strength of superconducting pairing within and between the two bands. The coefficients are defined as $a_0 = 2w/\lambda_0$, $a_1 = 1 + \lambda_+/\lambda_0$, and $a_2 = 1 - \lambda_-/\lambda_0$, where $w = \lambda_{11}\lambda_{22} - \lambda_{12}\lambda_{21}$, $\lambda_0 = (\lambda_-^2 + 4\lambda_{12}\lambda_{21})^{1/2}$, and $\lambda_- = \lambda_{11} - \lambda_{22}$.

From the fits to $\mu_0 H_{c2}(T)$ the diffusivity ratio is found to be $\eta = 0.20$ for I // a and $\eta = 0.124$ for I // b, indicating substantial disparity in electronic mobilities between the two bands responsible for the coexistence and competition between the CDW and SC collective electronic orders. Such imbalance naturally explains the positive curvature of $\mu_0 H_{c2}(T)$ and provides further evidence that superconductivity in $Hf_{0.02}Zr_{0.98}Te_3$ emerges from the interplay of multiple electronic bands, likely linked to the coexistence and competition between CDW and superconducting collective electronic states.

It is important to notice that the intraband and interband coupling constants ($\lambda_{11}$, $\lambda_{22}$, and $\lambda_{12} = \lambda_{21}$, see inset of figures 4c and 4d) yielded by the fits are significantly smaller for I ∥ a than for I ∥ b. This reduction is consistent with the presence of a charge density wave (CDW) instability around 53 K along the a-axis, which partially gaps the Fermi surface and weakens superconductivity in that direction. The excellent agreement between the $\mu_0 H_{c2}(T)$ data and the two-band model across the full temperature range provides strong evidence that superconductivity in $Hf_{0.02}Zr_{0.98}Te_3$ is unconventional, and intimately linked to the multiband electronic structure, and to the interplay between CDW and SC.

To explore further the superconducting behavior of $Hf_{0.02}Zr_{0.98}Te_3$ it is important to analyze the magnetization curves just below $T_c$, as shown in Fig. 5a. The Meissner shielding at 1.9 K is rapidly suppressed at very low fields (≈ 0.20 mT), indicating that magnetic flux starts to penetrate the sample. As the field increases further, a clear crossover from the linear Meissner response to the mixed (vortex) state is observed. This behavior confirms that the compound is a type-II superconductor, consistently with the $\rho(T)$ data discussed previously.

The lower critical field $H_{c1}(T)$ was extracted from the isothermal magnetization curves (Fig. 5a) by identifying the magnetic field at which the magnetization deviates from the initial linear Meissner region. To identify this point in a consistent and reproducible way, we adopted a fixed deviation criterion of $\Delta M = 10^{-3}$ emu/cm³ to mark the onset of vortex penetration, following the procedure described in Refs. [29–31]. This method ensures that $H_{c1}$ is determined from the first detectable departure from ideal diamagnetism.

The behavior of $H_{c1}(T)$ shown in Fig. 5b was fit using a single-band model typically applied to conventional BCS superconductors. However, even within the limited temperature range accessible in our measurements ($T \geq 1.9$ K), the experimental data cannot be satisfactorily described by this single-band approach. The systematic deviation between the model and the data suggests that the superconducting state cannot be accounted for by a single isotropic gap, giving credence to the possibility of multiple electronic bands contributing to superconductivity. This conclusion is fully consistent with the multiband behavior previously inferred from the analysis of the upper critical field in Figs. 4c and 4d.

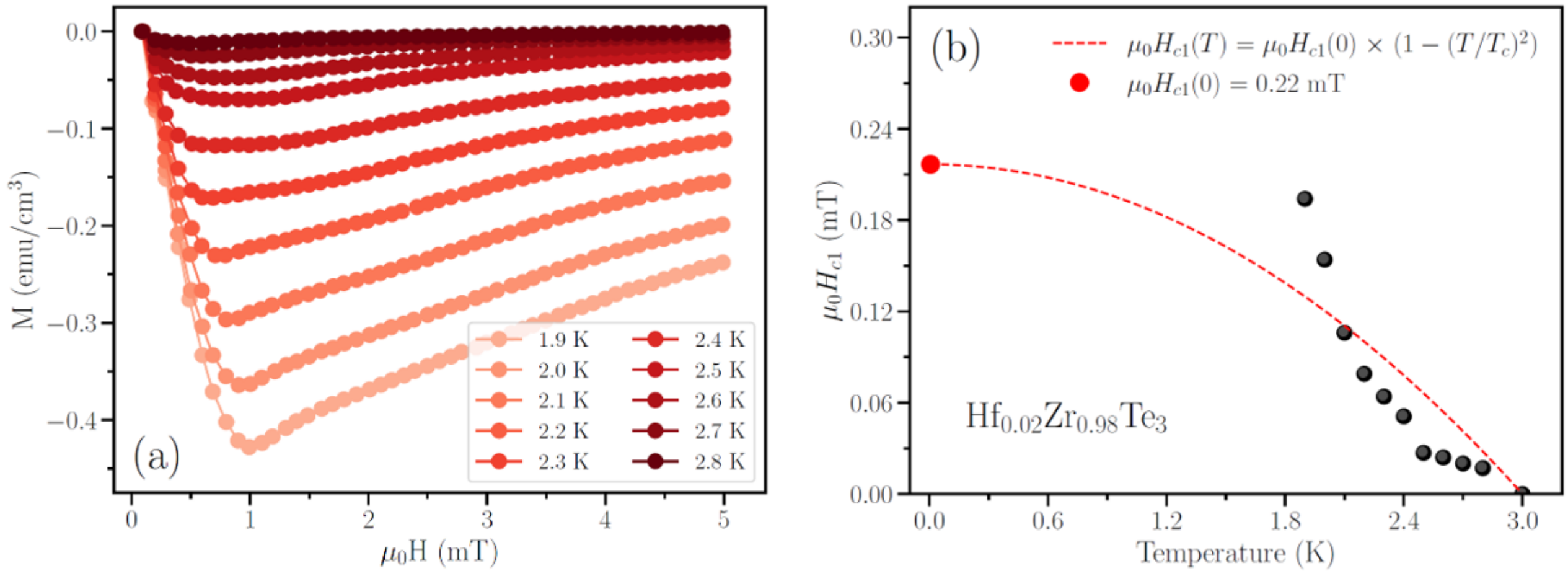

**Figure 5.** (a) Isothermal magnetization curves M vs $\mu_0H$ for $Hf_{0.02}Zr_{0.98}Te_3$ in the 1.9 - 2.8 K range. The linearity at low fields is disrupted by vortex penetration at $\mu_0H_{c1}$; (b) Temperature dependence $\mu_0H_{c1}$ extracted from the magnetization data and fit within a single-band model. The extrapolated zero-temperature value obtained from the fit is $\mu_0H_{c1}(0) \approx 0.22$ mT.

The charge carrier concentration $n$ for the $Hf_{0.02}Zr_{0.98}Te_3$ single crystal was estimated from the Hall coefficient using the approximation $R_H = 1/nq$, where q is the elementary charge. The results for both crystallographic configurations are presented in Fig. 6a (I // a) and Fig. 6b (I // b). The insets reveal that in the low-temperature range 1.8 - 20 K the carrier concentration remains nearly independent of temperature, with values of approximately $1.08 \times 10^{27}$ m$^{-3}$ for I // a and $2.71 \times 10^{27}$ m$^{-3}$ for I // b. This marked difference reflects the pronounced electronic anisotropy with a significantly lower carrier density along the a-axis. A clear change in the curvature of $R_H$ (T) is observed near 53 K, accompanied by an abrupt increase in the Hall resistance. This feature correlated well in temperature with the onset of the CDW transition, and it is consistent with the partial gapping of the Fermi surface.

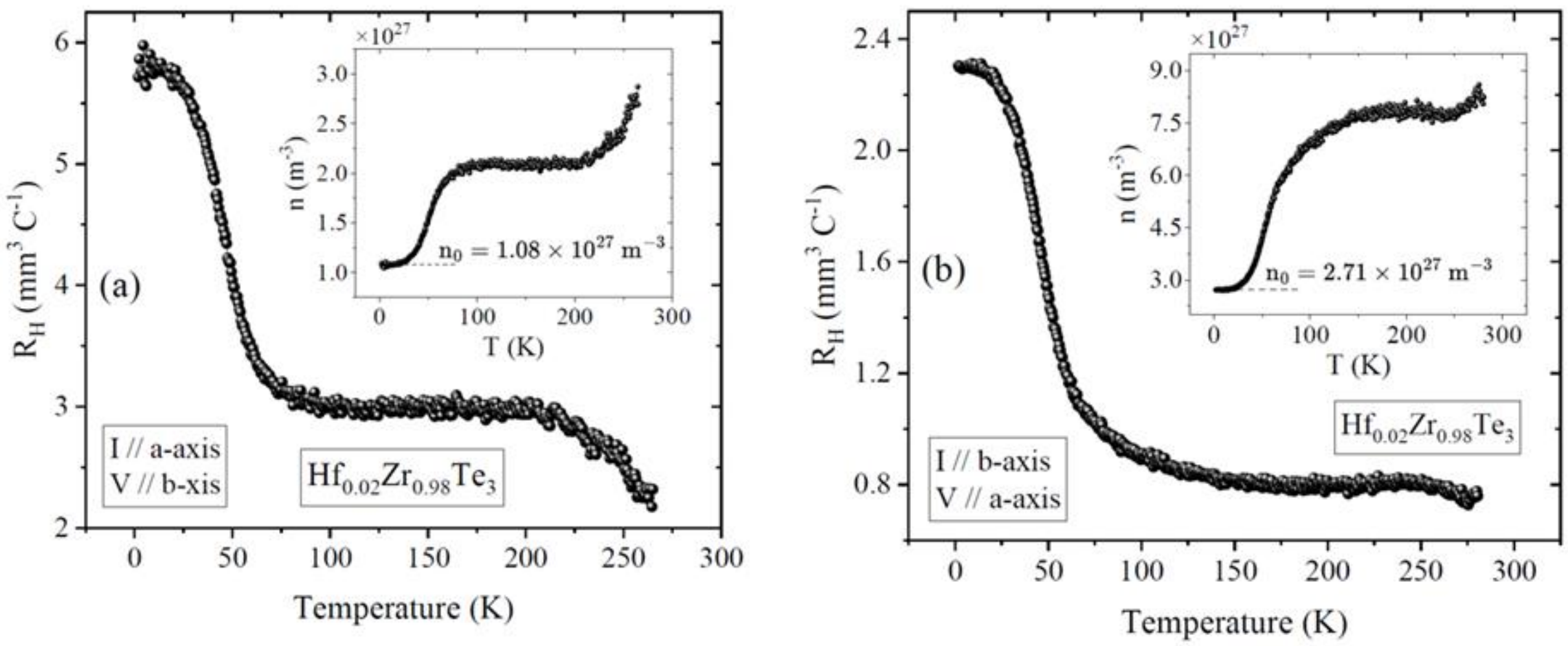

**Figure 6.** (a) Hall resistance of $Hf_{0.02}Zr_{0.98}Te_3$ measured with excitation current applied along the a-axis (I // a) and Hall voltage detected along the b-axis. The inset shows the temperature dependence of the corresponding charge carrier concentration; (b) Hall resistance measured with current applied along to the b-axis (I // b) and Hall voltage detected along the a-axis, with the inset displaying the carrier concentration as a function of temperature. Hall measurements were carried out in magnetic fields of +9 T and −9 T, with a 5 mA excitation current.

Measurements of specific heat and Seebeck coefficient as a function of temperature for $Hf_{0.02}Zr_{0.98}Te_3$ are presented in Fig. 7, and they show that the thermal behavior is consistent with the findings of the electrical and magnetic properties. The normal-state specific heat at low temperatures (Fig. 7a) was analyzed within the Sommerfeld–Debye framework, using the relation $C_p/T = \gamma + \beta T^2$, where $\gamma$ and $\beta$ correspond to the electronic and phonon contributions, respectively. A linear fit in the $T^2$ range between 10 and 100 $K^2$ yields $\gamma = 3.46$ mJ $mol^{-1}$ $K^{-2}$ and $\beta = 1.30$ mJ $mol^{-1}$ $K^{-4}$, revealing a significant electronic contribution at low temperatures. A clear anomaly is observed near 3.0 K (insets), in excellent agreement with the superconducting transition identified from resistivity and magnetic susceptibility measurements.

The Debye temperature was estimated using $\theta_D = \left(\frac{12\pi^4 NR}{5\beta}\right)$, where N is the number of atoms per formula unit (N = 4 for $ZrTe_3$) and R is the gas constant, yielding $\theta_D \approx 181.3$ K. The normalized specific heat anomaly at $T_c$ estimated from these data is $\Delta C_e/\gamma T_c \approx 0.43$ (lower inset 7b), about 30% of the BCS weak-coupling value of 1.43, suggesting weak electron–phonon coupling. The electron–phonon coupling constant $\lambda$ can be estimated from McMillan's expression [32],

$$\lambda = \frac{1.04 + \mu \log\left(\frac{\theta_D}{1.45T_c}\right)}{1 - 0.62\mu \log\left(\frac{\theta_D}{1.45T_c}\right) - 1.04} \quad (2)$$

using $\mu^* = 0.13$ which yields $\lambda \approx 0.59$ [11, 14, 15]. These values of $\Delta C_e/\gamma T_c$ and $\lambda$ for $Hf_{0.02}Zr_{0.98}Te_3$ are within the range found in weakly coupled superconductor.

The Seebeck coefficient (S) as a function of temperature is displayed in Fig. 7b. This measurement was carried out with a temperature differential established along the b-axis. The thermoelectric potential is consistently positive above $T_c$ indicating that the transport properties are dominated by hole-like charge carriers. This behavior is consistent with the hole-dominated three-dimensional band reported for $ZrTe_3$ [10, 33, 34]. From 300 K down to the CDW transition temperature ($T_{CDW} \approx 53$ K), the Seebeck coefficient decreases almost linearly, with a slope $\partial S/\partial T \approx 0.025$ μV/K². A clear change in slope is observed near $T_{CDW}$, below which the slope increases nearly 10-fold to $\partial S/\partial T \approx 0.245$ μV/K². It is tempting to assign this drastic change in the $\partial S/\partial T$ slope to the partial gapping of the Fermi surface at $T_{CDW}$. The value of S goes towards zero below $T_c$, consistent with the onset of superconductivity, though the resolution of the apparatus (≈ 1μV) is not sufficient for a clear determination.

Within a simplified single-band approximation, the nearly linear temperature dependence of the Seebeck coefficient just above $T_c$ can be understood in terms of the diffusive contribution of a Fermi liquid. In this picture, the temperature derivative of S is inversely related to the Fermi temperature $T_F$ [35], providing a useful way to estimate key electronic parameters from thermoelectric data, given by

$$\frac{\partial S}{\partial T} = \pm\frac{\pi^2 k_B}{2eT_F} \tag{3}$$

Although the superconductivity in $ZrTe_3$-based materials is consistent with multiband participation, the simplified single-band approach still offers a useful and independent way to estimate the charge carrier density in $Hf_{0.02}Zr_{0.98}Te_3$, which can be compared to the values obtained from Hall effect measurements. Using $\partial S/\partial T \approx 0.245$V/K² from the inset of Fig. 7b, the value of $T_F$ obtained from Eq. 3 is $T_F \approx 1735.71$ K, and the charge carrier concentration can be estimated within the free-electron gas approximation using the relation [11, 35]

$$\gamma = \frac{\pi^2 k_B n}{2T_F} \tag{4}$$

The free-electron gas approximation yields a carrier concentration $n \approx 1.25 \times 10^{27}$ m⁻³, which is in reasonable agreement with the values obtained from the Hall resistance, $1.08 \times 10^{27}$ m⁻³ (I // a - Fig. 6a), but differs somewhat from the $2.71 \times 10^{27}$ m⁻³ value obtained for I // b, further highlighting the pronounced electronic anisotropy of the system.

Using the relation $k_B T_F = \hbar^2 k_F^2/2m^*$, with $k_F = (3\pi^2 n)^{1/3}$, and combining the Fermi temperature extracted from the Seebeck data with the carrier concentration obtained from the Hall measurements, the effective mass was estimated to be $m^* \approx 2.57m_0$ for I // a and $m^* \approx 4.74m_0$ for I // b, where $m_0$ is the free-electron mass. The orientation dependence of the effective mass is consistent with the anisotropic electronic structure.

Evidence from chemically modified systems such as $Ni_{0.05}ZrTe_3$ [14], $Cu_{0.05}ZrTe_3$ [15], and $ZrTe_{2.96}Se_{0.04}$ [5], suggest that lattice distortions and chemical substitutions significantly alter the nesting conditions of the quasi-1D Fermi surface, leading to CDW suppression and the emergence of superconductivity. In those compounds, the reported effective masses ($2.2m_0$, $2.8m_0$, and $1.53m_0$, respectively) remain relatively moderate, indicating that superconductivity develops primarily as the CDW gap weakens and portions of the Fermi surface are restored.

A contrasting trend is observed in $Tb_{0.02}ZrTe_3$ [11], where the effective mass reaches $\approx 3.97m_0$ and the highest $T_c$ among $ZrTe_3$-derived systems is reported. In that case, CDW suppression is accompanied by additional electronic correlations associated with the partially filled 4f shell of Tb, suggesting that enhanced many-body interactions, rather than CDW suppression alone, are a primary factor in increasing $T_c$.

In contrast, Hf substitution is isovalent and does not introduce additional carriers or local magnetic moments. Instead, it primarily induces subtle structural perturbations while preserving the underlying quasi-1D electronic topology. The relatively moderate mass along the chain direction ($m^* \approx 2.57m_0$ for $I // a$) is consistent with the persistence of a robust CDW transition near 53 K. However, the markedly enhanced transverse mass ($m^* \approx 4.74m_0$ for $I // b$) points to strong anisotropic renormalization effects, likely reflecting the interplay between interchain coupling and CDW-driven partial gapping of the Fermi surface.

The superconductivity in $Hf_{0.02}Zr_{0.98}Te_3$ emerges within a regime where the CDW is robust, and with an effective mass that depends strongly on the direction. Therefore, unlike Tb-intercalated materials, where the CDW suppression and enhanced correlations cooperate to raise $T_c$, Hf substitution appears to strengthen the CDW ground state, while curbing superconductivity.

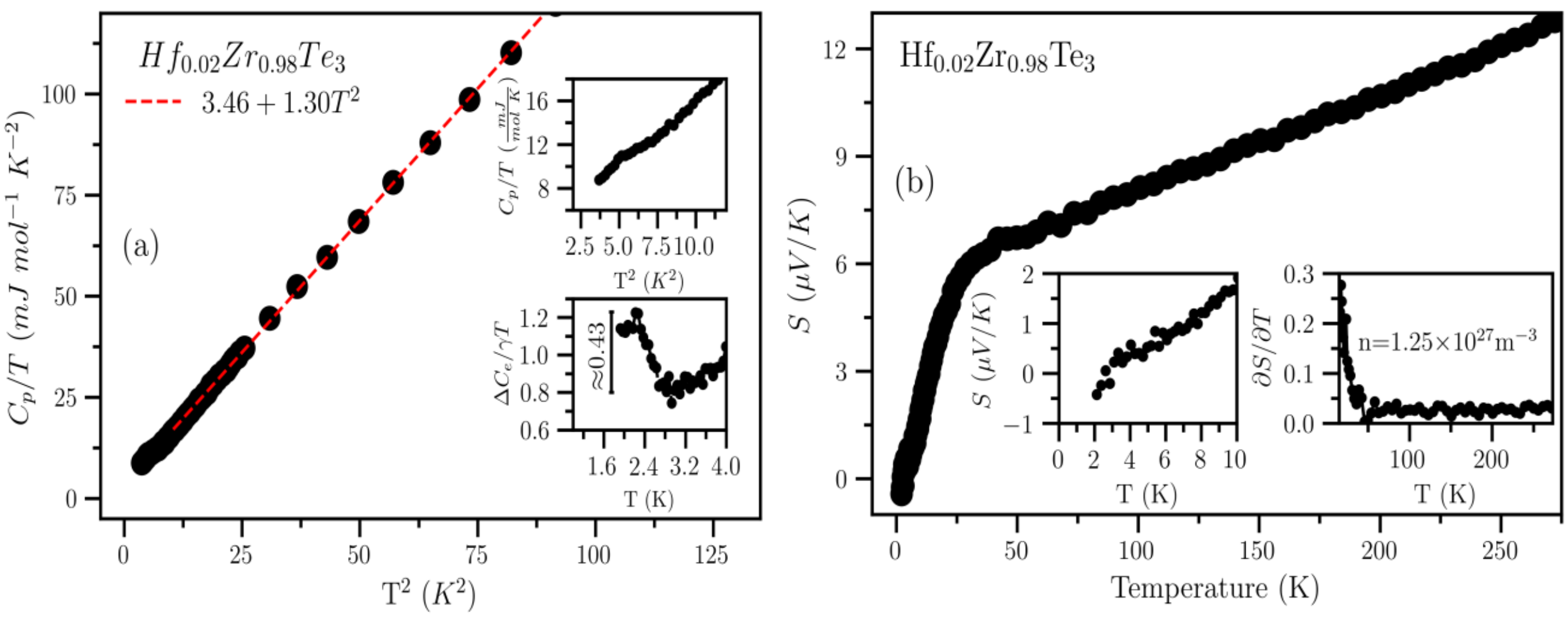

**Figure 7.** (a) $C_p/T$ versus $T^2$ plot of the low-temperature specific heat of $Hf_{0.02}Zr_{0.98}Te_3$. A fit of data to the Sommerfeld–Debye model expression $C_p/T = \gamma + \beta T^2$ yielded $\gamma$ and $\beta$ values of 3.16 $mJ.mole^{-1}.K^{-2}$ and 1.30 $mJ.mole^{-1}.K^{-4}$, respectively. The upper inset in the $T^2$ range from 3-12 $K^2$ range highlights a small anomaly due to the onset of superconductivity, while the lower inset shows a jump $\Delta C_e/\gamma T \approx 0.43$ corresponding to the onset of superconductivity at $T_c$; (b) Seebeck coefficient measured from 2 to 300 K, measured with a temperature gradient along the b-axis. The left inset shows S dropping to zero near 3.3 K, consistent with superconductivity, while the right inset (dS/dT) reveals a slope change around 53 K, associated with the CDW transition.

The superconducting parameters for the $ZrTe_3$-based materials reported in the literature and the findings of this work on $Hf_xZr_{1-x}Te_3$ are summarized in Table 1.

Table I: Superconducting parameters of $ZrTe_3$-based materials reported in the literature.

| Parameter | $Hf_{0.02}Zr_{0.98}Te_3$ | $ZrTe_3$ | $Cu_{0.05}ZrTe_3$ | $Ni_{0.05}ZrTe_3$ | $ZrTe_{2.96}Se_{0.04}$ | $Tb_{0.02}ZrTe_3$ |
|---|---|---|---|---|---|---|
| $T_c$ (K) | 3.3 | 2.0 | 3.8 | 3.1 | 4.4 | 5.4 |
| $T_{CDW}$ (K) | 53 | 63 | 55 | 41 | – | 50 |
| $H_{c2}(0)$ (T) H∥c | 0.51 | – | 1.0 | 0.49 | 1.25 | 1.5 |
| $\lambda$ | 0.59 | – | 0.68 | 0.63 | – | 0.79 |
| $\theta_D$ (K) | 181.3 | 199 | 186 | 192.4 | 184 | 165 |
| $\gamma\ mJ.mol^{-1}.$ | 3.46 | 2.23 | 2.64 | 2.7 | 3.4 | 3.3 |
| $m/m_0$ | 2.57 | – | 2.2 | 2.8 | 1.53 | 3.97 |
| n ($10^{27}m^{-3}$) | 1.08 | – | 1.30 | 0.65 | – | 1.15 |
| Ref. | This work | [12] | [15] | [14] | [5] | [11] |

In order to probe the evolution of the CDW and superconductivity as a function of pressure, we carried out measurements of resistivity in pressures up to ≈ 2 GPa, with excitation current applied along the a-axis. As displayed in Fig. 8a, pressure drives $T_{CDW}$ up from ≈ 53 K at ambient pressure to ≈ 92 K at 1.69 GPa, while superconductivity is suppressed at a fast rate of about ≈ -4.3 K/GPa. The opposite trends of $T_{CDW}$ and $T_c$ under pressure is consistent with the competition of these two ground states for the Fermi surface. The room temperature resistivity drops by ≈ 30% from ambient pressure to 1.88 GPa. A *T–P* phase diagram mapping the CDW and superconducting phase is show in Fig. 8b.

It is plausible to consider that lattice compression yielded by pressure and the chemical pressure caused by the Hf partial substitution enhance the orbital overlap along the quasi-one-dimensional Te chains, therefore strengthening Fermi-surface nesting and stabilizing the CDW state. As the CDW gap expands over a larger portion of the Fermi surface, the density of states available for Cooper pairing is reduced, naturally suppressing superconductivity. This suggests that Hf substitutions, just like pressure, can be used to fine-tune the competition between CDW and superconductivity, primarily through a modification of the electronic structure, and by causing an effective chemical pressure in the crystal lattice, reinforcing the quasi-one-dimensional instability of the system.

A similar pressure-driven increase in $T_{CDW}$ has been reported for pristine $ZrTe_3$ [13, 18]. In the pure $ZrTe_3$ case, $T_{CDW}$ initially increases with pressure, peaks near ≈ 2 GPa, and starts to drop again, until the CDW is suppressed abruptly near 5 GPa. [13, 18]. The suppression of the CDW near 5 GPa coincides with the reemergence of superconductivity, with $T_{c,\ onset}$ reaching a maximum value (≈ 6 -7 K) at the highest value of the experiments (≈ 10 - 20 GPa) [18, 24].

The effect of pressure on the competition of CDW and superconductivity in $Hf_{0.02}Zr_{0.98}Te_3$ appears to arise from subtle modification in the electronic structure, rather than structural modifications. Strengthening of the 1D component of the Fermi surface seems to favor CDW formation. The lack of a CDW feature in $\rho_b(T)$ suggests the Fermi surface is not too affected by gap opening along the 1D density of states, a point that needs to be further investigated [12, 13, 18, 25]. Band-structure calculations suggest that the pressure-induced reinforcement of the CDW state and simultaneous suppression of superconductivity cannot be explained solely by a simple nesting scenario, underscoring the complex interplay between these two collective states [25, 26]

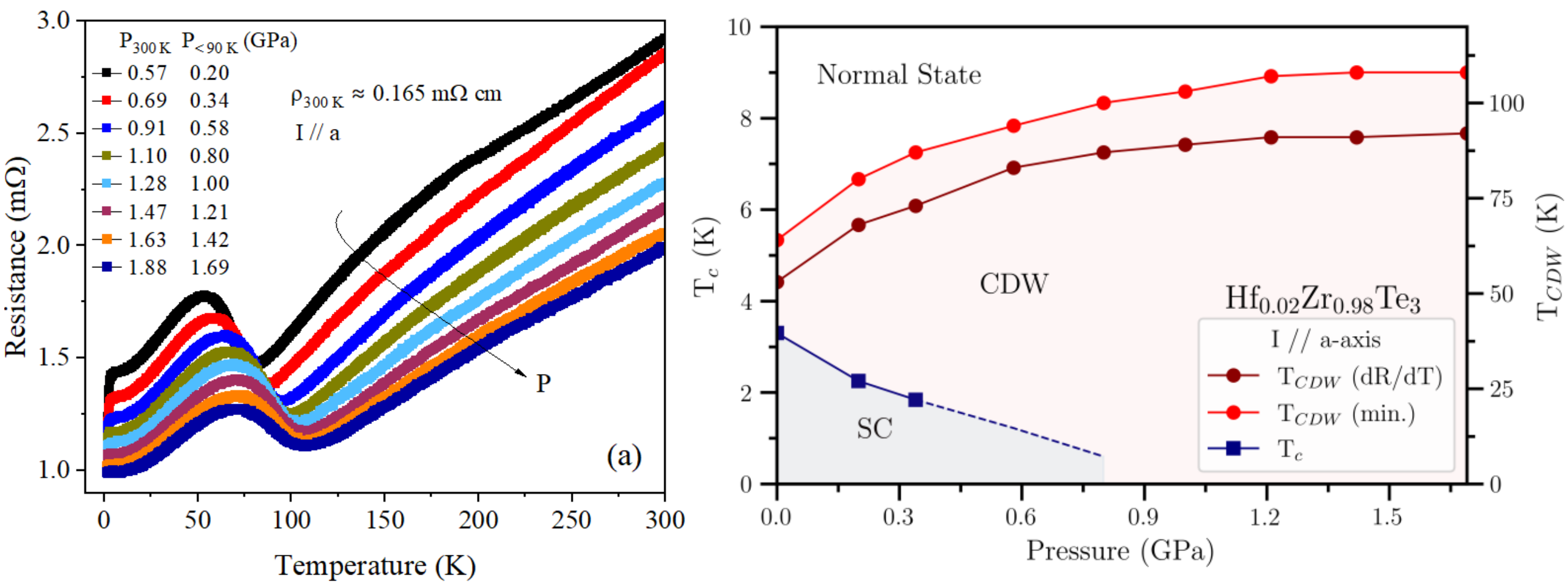

**Figure 8.** (a) Temperature dependence of the electrical resistivity $\rho_a(T)$ for $Hf_{0.02}Zr_{0.98}Te_3$ in pressures up to 1.9 GPa; (b) Pressure–temperature (T–P) phase diagram for $Hf_{0.02}Zr_{0.98}Te_3$. The superconducting phase is delimited by the transition temperature $T_c$ (left y-axis), determined from the $\rho_a(T)$ curves using a 10% resistive drop criterion. For P > 0.5 GPa $T_c$ was determined by extrapolation. The charge density wave (CDW) phase is delimited using two methods: (i) the inflection-point, where $T_{CDW}$ (right y-axis) is taken from the minimum in dR/dT; and (ii) the onset, where $T_{CDW}$ is identified from the minimum in $\rho_a(T)$.

## Conclusions

This study of the electrical, magnetic and thermal properties of the low dimensional $Hf_{0.02}Zr_{0.98}Te_3$ shows that a slight isovalent partial substitution of Hf for Zr is sufficient to stabilize superconductivity, and to affect the competition between superconductivity and CDW order. While superconductivity in undoped $ZrTe_3$ with $T_c$ ≈ 2.0 K can be characterized as filamentary, superconductivity in $Hf_{0.02}Zr_{0.98}Te_3$ with a $T_{c,\ onset}$ ≈ 3.3 K is a bulk phenomenon, as indicated by the specific heat and magnetization measurements. The value of $T_{CDW}$ drops from ≈ 63 K in pure $ZrTe_3$ to ≈ 53 K in $Hf_{0.02}Zr_{0.98}Te_3$, suggesting that even at this minor level of substitution, the CDW gaps a smaller portion of Fermi surface, leaving a larger density of states to sustain superconductivity at lower temperatures.

The temperature dependence of the $H_{c1}$ and $H_{c2}$ critical fields for $Hf_{0.02}Zr_{0.98}Te_3$ deviates markedly from single-band behavior. An attempt to fit these critical field data to a multiband model was quite successful. The pronounced anisotropy of the superconducting parameters yielded in these fits clearly reflects the strong anisotropy.

Measurements of the electrical resistivity in hydrostatic pressure in $Hf_{0.02}Zr_{0.98}Te_3$ showed an increase in $T_{CDW}$ and a drop in $T_c$, behaviors consistent with the competition of superconductivity and CDW. Pressure raises $T_{CDW, inflection}$ from ≈ 53 K (P=0) to ≈ 92 K (1.69 GPa) while suppressing $T_c$ rapidly, at the rate of ≈ -4.3 K/GPa. Given that $T_{CDW}$ also increases with the Hf content (Fig. 2), the observed increase in $T_{CDW}$ with pressure is quite possibly the result of the combined effect of chemical and thermodynamic pressures. Lattice compressions are likely to reinforce the one-dimensional electronic instability responsible for the CDW in this pressure range, rather than strengthening the three-dimensional electronic coherence required for superconductivity.

In summary, the detailed electrical, magnetic and thermal behavior of $Hf_{0.02}Zr_{0.98}Te_3$ suggest that partial substitutions as well as pressure can be used as fine-tuning parameters to affect Fermi-surface nesting, CDW order, and multiband superconductivity in $ZrTe_3$-based systems. Further studies at higher Hf concentrations and pressures are in order.

## ACKNOWLEDGEMENTS

The authors gratefully acknowledge São Paulo Research Foundation (FAPESP) under Grants No. 2024/01514-2, 2024/23535-1, FAPEMIG (APQ-05247-23) for the financial support of this research.